\documentstyle[epsf,amstex,amssymb,righttag,12pt]{article}

\newcommand{\sres}{\operatorname{sres}}

\begin{document}
\title{\Large\sc On the Integrable Hierarchies \\
Associated With
$N=2$ Super $W_n$ Algebra}
\author{Q. P. Liu\thanks{On leave of absence from
Beijing Graduate School, CUMT, Beijing 100083, China}
\thanks{Supported by {\em Beca para estancias temporales
de doctores y tecn\'ologos extranjeros en
Espa\~na: SB95-A01722297}}
\\Departamento de F\'\i sica Te\'orica, \\
Universidad Complutense,\\ 
E28040-Madrid, Spain.}

\date{}

\maketitle

\begin{abstract}
A new Lax operator
is proposed from the viewpoint of constructing
the integrable hierarchies related with $N=2$ super
$W_n$ algebra. It is shown that the Poisson algebra
associated to the second Hamiltonian
structure for the resulted hierarchy contains
the $N=2$ super Virasoro algebra as a proper
subalgebra. The simplest cases are discussed in detail.
In particular, it is proved that the supersymmetric two-boson hierarchy is
one of $N=2$ supersymmetric KdV hierarchies. Also, a Lax operator
is supplied for one of $N=2$ supersymmetric Boussinesq
hierarchies.

\end{abstract}
\newpage
\section{Introduction}
Recently, $N=2$ supersymmetric integrable hierarchies
have attracted much attention(see \cite{b} for example).
 It is believed that
a full understanding of these hierarchies
will shed  light on the study of conformal
field theory. Besides their physical relevance, these
hierarchies are mathematically interesting. One of the 
appealing problems, which has not been explained,
is why there exist three different $N=2$ supersymmetric 
hierarchies.

One way to construct $N=2$ super $W_n$ algebra
is to use Gel'fand-Dickey bracket for an
odd order super differential operator $L$ below. For the cases
of $n=2, 3$ and $4 $, the resulted brackets are indeed
the corresponding
$N=2$ super $W_n$ algebras respectively\cite{fr}-\cite{ik}, 
although the situation in general is still a conjecture.
We will argue that two
Lax operators can be formed out of $L$. Among them, one is
the one proposed by Inami and Kanno\cite{ik}\cite{ik1}, the other one
is new in general. For the new proposed one, our calculation
shows that the Poisson algebra associated to the second
Hamiltonian structure contains $N=2$ super Virasoro
algebra in proper. This is the evidence that the general
case will lead to $N=2$ super $W_n$ algebra. We will
further show that this is indeed the case
 in the simplest examples.

The layout of the letter is as follows. We consider 
the general case in next section and calculate explicitly
the $N=2$ super Virasoro algebra. Section three is
devoted to the simplest examples. In particular, we
will show that the supersymmetric two-boson(or Kaup-Broer)
 hierarchy of Brunelli and Das\cite{bd}
is equivalent to one of the $N=2$ supersymmetric KdV
hierarchies
of Laberge and Matheiu\cite{lm}.
In the next simplest case, our system
is identical to so-called $N=2$ supersymmetric Boussinesq
equation whose Lax pair was not known before, so we provide
a Lax operator for this system in $N=1$ form.

\section{$N=2$ Super $W_n$ Algebra and Lax Operators}
The general $N=2$ super $W_n$ algebra is believed to be 
related with an odd order super differential operator\cite{fr}-\cite{ik}
\begin{equation}
L=D^{2n-1} +u_{2n-3}D^{2n-3}+\dots +u_0,
\end{equation}
where $D={\partial\over\partial {\vartheta}}+\vartheta{\partial\over\partial x}$ 
is the 
super derivative. Indeed, given the operator $L$, one may calculate
the second Gel'fand-Dickey bracket from the Poisson tensor
\begin{equation}\label{gd}
\Theta :\qquad \Delta H\longrightarrow (L\Delta H)_{\geq 0}L-
L(\Delta H L)_{\geq 0}, 
\end{equation}
where $\Delta H$ is properly parametrized(see \cite{fr} for example).
We also use the standard notations for a given super 
pseudo-differential operator: 

$
A=\sum_{i\geq 0}a_iD^i+
\sum_{i\leq -1}D^ia_i:=\sum_{i\geq 0}A_i+\sum_{i\leq -1}A_i,
\qquad \sres A=a_{-1}.
$
\smallskip

Since the operator $L$ is lacking of term $D^{2n-2}$, we
have to modify the general tensor(\ref{gd}) 
so that it will give us a 
proper expression. The situation is much same as in the pure
bosonic case and the reduced Poisson tensor is
\[
{\hat{P}} : \Delta H\longrightarrow (L\Delta H)_{\geq 0}L-L(\Delta H L)_{\geq 
0}+(-1)^{\mid\Delta H\mid}
[ L,{\cal D}^{-1}\textstyle{\sres}[L,\Delta H] ],
\]
where ${\cal D}^{-1}=\int^z dxd\vartheta $ denotes the super integration
with $z=(x,\vartheta)$
and $[,]$ means the graded commutator. The parity of $\Delta H$ is
indicated by $\vert \Delta H\vert$.

That this approach indeed supplies the $N=2$ super $W_n$ algebra is
worked out explicitly for the simplest three cases
$n=2$, $n=3$ and $n=4$ in \cite{fr}-\cite{ik}\cite{yw},
while the correctness of 
the general case is still an unproved conjecture.

Let us consider the problem of constructing
integrable hierarchies associated with this algebra.
Since the $L$ is an odd operator, it is not possible
to construct integrable hierarchies directly with it.
To have meaningful results, let us
define two Lax operators via $L$
\begin{equation}
\begin{aligned}
L_{1}^{(n)}=&\partial^{n}+v_{2n-2}D^{2n-2}+v_{2n-3}D^{2n-3}+\dots+v_1 D,\\
L_{2}^{(n-1)}=&\partial^{n-1}+w_{2n-4}D^{2n-4}+w_{2n-5}D^{2n-5}+\dots+w_0+D^{-1}
w_{-1},
\end{aligned}
\end{equation}
and $L_{1}^{(n)}$ and $L_{2}^{(n-1)}$ are related to $L$ by
\begin{equation}\label{re}
L_{1}^{(n)}=LD,\qquad L_{2}^{(n-1)}=D^{-1}L.
\end{equation}

We notice the $L_{1}^{(n)}$ is proposed first by Inami and Kanno
in the context of $W$ algebra\cite{ik}\cite{ik1}, 
there it is shown that this Lax
operator gives one of $N=2$ supersymmetric hierarchies.

We concern here with the operator $L_{2}^{(n-1)}$. Since
 the $L_{2}^{(n-1)}$ is
of type of constrained Modified KP\cite{os}, we may call
it constrained super modified KP. The integrable hierarchy
can be constructed by means of standard fractional power
approach
\begin{equation}\label{hi}
{\partial \over\partial t_{k}}L_{2}^{(n-1)}=
[((L_2^{(n-1)})^{k\over2})_{\geq 1},L_{2}^{(n-1)}],
\end{equation}
as for the Hamiltonian structures, we
may use the results of Oevel and Strampp\cite{os}\cite{o}.
The system(\ref{hi}) is indeed a bi-Hamiltonian system with
the first Poisson tensor given by
\begin{equation}\label{o}
Q: \Delta H\longrightarrow ([L_{2}^{(n-1)},\Delta H])_{\geq -1},
\end{equation}
and the second one reads as
\begin{equation}\label{os}
\begin{aligned}
P: \Delta H\longrightarrow &(L_{2}^{(n-1)}\Delta H)_{\geq 0})L_{2}^{(n-1)}-
L_{2}^{(n-1)}(\Delta H L_{2}^{(n-1)})_{\geq 0}\\
&+
\left[L_{2}^{(n-1)}, (L_{2}^{(n-1)}\Delta H)_0\right]
-D^{-1}\big(\sres[\Delta H,L_{2}^{(n-1)}]\big)L_{2}^{(n-1)}\\
&+\big[{\cal D}^{-1}\big(\sres[\Delta H,L_{2}^{(n-1)}]\big),
L_{2}^{(n-1)}\big],
\end{aligned}
\end{equation}
where $\Delta H$ is parametrized as
\[
\Delta H={\delta H\over \delta w_{-1}}-\sum_{i=0}^{2n-4}(-D)^{-i-1}
{\delta H\over \delta w_i},
\]
with ${\delta H\over \delta w_{2i-1}}$ are bosonic and the rest
of them fermionic.
\smallskip

Let us now calculate the subalgebra for the first two
coefficients. To this end, we take $\Delta H$ as 
\[
\Delta H=\partial^{-n+2}(D^{-1}\Lambda+X),\qquad
\Lambda={\delta H\over\delta w_{2n-4}},\qquad
X=-{\delta H\over\delta w_{2n-5}},
\]
calculating (\ref{os})
and picking up the coefficients of $\partial^{n-2}$ and $\partial^{n-3}D$,
we have
\begin{equation}\label{vir}
\begin{aligned}
\{w_{2n-4}(z),w_{2n-4}(z')\}&=\big(n(n-1)\partial 
D+(Dw_{2n-4})-2w_{2n-5}\big)\Delta(z-z'),\\
\{w_{2n-4}(z),w_{2n-5}(z')\}&=-\left({n(n-1)\over2}\partial^2+\partial 
w_{2n-4}-w_{2n-5} D\right)\Delta(z-z'),\\
\{w_{2n-5}(z),w_{2n-5}(z')\}&=-\left(w_{2n-5}\partial+\partial 
w_{2n-5}\right)\Delta(z-z'),
\end{aligned}
\end{equation}
where $\Delta(z-z')$ is the super delta function. To see the connection
with the $N=2$ super Virasoro algebra, we perform 
the following invertible transformation
\[
J=u,\quad T=-\alpha+\frac{1}{2}(Du),
\]
brings the algebra (\ref{vir}) to 
\begin{equation}\label{vir1}
\begin{aligned}
\{J(z), J(z')\}&=(n(n-1)\partial D+2T)\Delta(z-z'),\\
\{J(z), T(z')\}&=(\partial J-\frac{1}{2}(DJ)D)\Delta(z-z'),\\
\{T(z), T(z')\}&=\left(\frac{1}{4}n(n-1)\partial^2D+\frac{3}{2}T\partial
+\frac{1}{2}(DT)D+T_x\right)\Delta(z-z'),
\end{aligned}
\end{equation}
which is nothing but the the $N=2$ super Virasoro algebra in the
$N=1$ form (cf.\cite{ik}\cite{hn}\cite{fr}). Thus, we see that the Poisson 
algebra
related with the Hamiltonian structure has the $N=2$ super Virasoro
algebra as its a proper subalgebra. This observation leads
us to a conjecture: {\em the hierarchy constructed out 
of $L_{2}^{(n-1)}$ is a hierarchy coincident with one
of the hierarchies associated with $N=2$ super $W_n$
algebra}.

{\em Remarks}: 
\begin{itemize}
\item The general formulae(\ref{vir}) are not valid in the
simplest case $n=2$. However we will see, in the next section,
we still have $N=2$ super Virasoro algebra;
\item The validity of our conjecture will be shown
 in the cases with $n=2$ and $n=3$ by
direct calculation.
\end{itemize}

\section{Examples}
We perform calculation in this section in the simplest
cases and show that our conjecture made above is indeed
valid in these concrete cases. For clarity, we will
 use the fields without index in Lax operators in the sequel.

\subsection{KdV or Kaup-Broer Case }

In the case $n=2$, we have 
\begin{equation}
L_{2}^{(1)}=\partial+u+D^{-1}\alpha,
\end{equation}
and the simplest flow($t_2$) is
\begin{equation}\label{t2}
u_t= u_{xx}+u^2+2(D\alpha)_x,\qquad
\alpha_t=-\alpha_{xx}+2(\alpha u)_x.
\end{equation}

We remark here that the Lax operator $L_{2}^{(1)}$ is essentially
the one proposed in\cite{bd} for the supersymmetric two-boson(or Kaup-Broer)
system. We
will see that as a by-product of our analysis, supersymmetric two-boson 
hierarchy
is equivalent to one of $N=2$ supersymmetric KdV hierarchy.
 
The first a few Hamiltonians are
\[
H_1=\int\alpha \, dz,\qquad H_2=\int u\alpha \, dz,\qquad 
H_3=\int\alpha(u_x+u^2+(D\alpha)) \,dz,
\]
two Poisson tensors are easily calculated
\begin{equation}
Q=\left(\begin{array}{cc}0&\partial\\
\partial&0\end{array}\right),\qquad
P=\left(\begin{array}{cc}2D\partial+2\alpha+(Du)&\partial^2
+\partial u+\alpha D\\
-\partial^2+u\partial -D\alpha&\alpha\partial+\partial\alpha
\end{array}\right),
\end{equation}
and the whole hierarchy is bi-Hamiltonian, in particular
the $t_2$ flow(\ref{t2}) can be written as
\[
\left(\begin{array}{c}u\\ \alpha\end{array}\right)_t=P
\left(\begin{array}{c}{\delta H_2\over\delta u}\\[1mm] 
{\delta H_2\over\delta\alpha}\end{array}\right)
 =Q\left(\begin{array}{c}{\delta H_3\over\delta u}\\[1mm] 
{\delta H_3\over\delta\alpha}\end{array}\right).
\]

By means of $L_{2}^{(1)}=D^{-1}L$ or
\[
\partial+u+D^{-1}\alpha=D^{-1}(D\partial +{\hat u}D+{\hat\alpha})
\]
we obtain
\begin{equation}\label{m0}
{\hat u}=u, \qquad {\hat \alpha}=\alpha +(Du).
\end{equation}

On the other hand, we may have a Poisson algebra
from the Gel'fand-Dickey bracket for $L$, which
 in terms of Poisson tensor reads as
\begin{equation}
{\hat P}=\left(\begin{array}{cc}2D\partial+2{\hat\alpha}
-(D{\hat u})&-\partial^2
+\partial {\hat u}-{\hat\alpha} D\\
\partial^2+{\hat u}\partial +D{\hat\alpha}&
{\hat\alpha}\partial+\partial{\hat\alpha}
\end{array}\right),
\end{equation}
a simple calculation shows that the invertible map(\ref{m0})
is a Hamiltonian map between $P$ and ${\hat P}$.

It is well known that the following map
\begin{equation}\label{m00}
J={\hat u},\qquad T={\hat \alpha}-{1\over 2}(D{\hat u}),
\end{equation}
converts the Poisson algebra defined by ${\hat P}$ to
the $N=2$ super Virasoro algebra with $n=2$ in (\ref{vir1}), so to bring
the Poisson algebra inherited from $P$ to the $N=2$
super Virasoro algebra, we need only compose the map(\ref{m0})
with (\ref{m00}). The composed map will bring
our $Q$ and $P$ to two Hamiltonian operators of
the $N=2$ supersymmetric KdV system with $a=4$\cite{lm}(see 
\cite{mp} also). This argument leads us to the conclusion
that the supersymmetric two-boson hierarchy\cite{bd}
is equivalent to the $N=2$, $a=4$ supersymmetric KdV
hierarchy.

\subsection{Boussinesq Case}

Now we work with the following Lax operator
\begin{equation}\label{lax}
L_{2}^{(2)}=\partial^2+u\partial+\alpha D +v+D^{-1} \beta,
\end{equation}
and the simplest flow($t_2$) reads as
\begin{equation}\label{t22}
\begin{aligned}
u_t&=2v_x,\qquad  \alpha_t=-2\beta_x,\\
v_t&=v_{xx}+2(D\beta)_x+uv_x+\beta (Du)+\alpha (Dv)+
2\alpha\beta,\\
\beta_t&=-\beta_{xx}+(\beta u)_x+D(\beta\alpha).
\end{aligned}
\end{equation}

The first a few Hamiltonians are
\begin{align*}
H_1&=-\int\alpha \,dz,  \qquad H_2=\int\beta\,dz,\\
H_3&={1\over 2}\int(\beta u-v\alpha
-{1\over 2}u\alpha_x+{1\over 4}\alpha(D\alpha)
+{1\over 4}u^2\alpha)\, dz,\quad
H_4=\int v\beta\,dz.
\end{align*}

We now calculate the Hamiltonian structures for the
related hierarchy. The first one is 
\[
Q=\left(\begin{array}{cccc}
0&0&0&2\partial\\
0&0&-2\partial&0\\
0&-2\partial&2\partial D-(Du)+2\alpha&\partial^2+u\partial+
\alpha D\\
2\partial&0&-\partial^2+\partial u-D\alpha&0\end{array}
\right),
\]
as for the second one, it is in rather complicated form
and for clarity, we present it in the table 1 below:
\newpage
\bigskip
\noindent{\sc Table 1:} Matrix entries of operator $P$.
\vspace*{-.29cm}

\hspace*{-.6cm}\rule{1\textwidth}{0.2 mm}\\ 
\vspace*{-.94cm}
   
\hspace*{-.6cm}\rule{1\textwidth}{0.15 mm} 
\vspace*{-1.2cm}

\begin{small}
\begin{align*}
P_{11}=& 6D\partial +(Du)-2\alpha, 
\hspace*{3.2cm} P_{12}= -3\partial^2 -\partial u+\alpha D,\\
P_{13}=& 4uD\partial +2(Du)\partial +4u_x D+2(Du)_x +(Dv)+2\beta,\\
P_{14}=& 2\partial^3+2\partial u\partial +2\partial\alpha D
        +2\partial v+\beta D,\\
P_{21}=& 3\partial^2 -u\partial -D\alpha,\hspace*{3.6cm}
P_{22}= -\alpha\partial -\partial\alpha,\\
P_{23}=& -2\partial^3+2\partial^2u-2\partial D\alpha-
2v\partial+D\beta -v_x,\\
P_{24}=& -3\beta \partial-2\beta_x,\hspace*{4cm}
P_{31}= 4u\partial D+2(Du)\partial +(Dv)+2\beta,\\
P_{32}=& -2\partial^3 -2u\partial^2-2\alpha \partial D
-2v\partial -\beta D-v_x,\\
P_{33}=& 2\partial^3 D -\partial^2(Du)-(Du)\partial^2+
2\alpha \partial^2+2\partial^2\alpha +v\partial D
+\partial Dv+Du\partial u\\&+u\partial uD+u(Dv)-v(Du)
+2u\beta+2\alpha v-\alpha u_x
-(D\alpha)(Du)+2\alpha (D\alpha),\\
P_{34}=& \partial^4+u\partial^3+\partial^2u\partial+\partial^2
\alpha D+\alpha\partial^2D+\partial^2v+u\partial u\partial
+u\partial\alpha D+\alpha Du\partial
-\beta\partial D\\
&+3(D\beta)\partial+u\partial v+\alpha(D\alpha)D
+\beta Du
+\alpha Dv-\beta_xD
+2\alpha\beta+2(D\beta)_x,\\
P_{41}=&2\partial^3-2\partial u\partial-2\alpha\partial D+
2(D\alpha)\partial+2v\partial-D\beta,\qquad
P_{42}= -2\beta\partial-\partial\beta,\\
P_{43}=& -\partial^4+\partial^3u+\partial u\partial^2
+\partial^2\alpha D+\alpha D\partial^2-\partial^2(D\alpha)-
(D\alpha)\partial^2
-\partial u\partial u
-v\partial^2\\
&+\partial uD\alpha-\alpha D\partial u+\beta\partial D+2(D\beta)\partial+
v\partial u
-vD\alpha +(D\alpha)\partial u+\beta uD\\
&-(D\alpha)D\alpha+(D\beta)_x
-(D\beta)u+2\beta\alpha+\alpha\alpha_x,\\
P_{44}=&\partial\beta u+\beta u\partial -2\beta_x\partial
+D\beta\alpha+\beta\alpha D-\beta_{xx},
\end{align*}
\end{small}

\vspace*{-1cm}

\hspace*{-.6cm}\rule{1\textwidth}{0.15 mm}\\ 
\vspace*{-.94cm}
   
\hspace*{-.6cm}\rule{1\textwidth}{0.2 mm}   

\bigskip
\smallskip

We notice that in particular the $t_2$-flow(\ref{t22})
 can be written
as
\[
{\boldsymbol u}_{t}=P{\delta H_2\over\delta {\boldsymbol u}}=
Q{\delta H_4\over\delta {\boldsymbol u}},
\]
where ${\boldsymbol u}=(u,\alpha,v,\beta)$.

As we conjectured in last section,
the Poisson algebra associated with $P$
should be $N=2$ super $W_3$ algebra. To justify this, 
we recall that the algebra
is constructed out of the operator
\begin{equation}
{\hat L}=\partial^2D-{\hat u}\partial D-{\hat\alpha}\partial
-{\hat v}D-{\hat \beta},
\end{equation}
using Gel'fand-Dickey second bracket. The above
form is chosen for the comparison with \cite{ik}.
The explicit expression is listed in the Table 2:

\bigskip
\noindent{\sc Table 2:} Matrix entries of operator ${\hat P}$.
\vspace*{-.29cm}

\hspace*{-.6cm}\rule{1\textwidth}{0.2 mm}\\ 
\vspace*{-.94cm}
   
\hspace*{-.6cm}\rule{1\textwidth}{0.15 mm}   

\vspace*{-.7cm}

\begin{small}
\begin{align*}
{\hat P}_{11}=& 6\partial D+(D{\hat u})-2{\hat\alpha},
\hspace*{4.2cm}
{\hat P}_{12}=-3\partial^2-\partial {\hat u}-{\hat\alpha} D,\\
{\hat P}_{13}=&-3\partial^2D-3\partial D{\hat u}+
\partial{\hat\alpha}+(D{\hat v})-2{\hat\beta},\\
{\hat P}_{14}=&2\partial^3+2\partial^2{\hat u}-2\partial D
{\hat\alpha}-2\partial {\hat v}+{\hat\beta} D,\\
{\hat P}_{21}=&3\partial^2-{\hat u}\partial-D{\hat\alpha},
\hspace*{4.6cm}
{\hat P}_{22}=-{\hat\alpha}\partial-\partial{\hat\alpha},\\
{\hat P}_{23}=&-\partial^3-\partial^2{\hat u}+
\partial D{\hat\alpha}-2{\hat v}\partial-{\hat v}_x-D{\hat \beta},
\hspace*{1cm}
{\hat P}_{24}= -3{\hat\beta}\partial-2{\hat\beta}_x,\\
{\hat P}_{31}=& 3\partial^2 D-3{\hat u}\partial D-{\hat \alpha}
\partial+(D{\hat v})-2{\hat \beta},\\
{\hat P}_{32}=&-\partial^3+{\hat u}\partial^2+{\hat\alpha}\partial D
-2{\hat v}\partial+{\hat \beta}D-{\hat v}_x,\\
{\hat P}_{33}=&-2\partial^3D-2\partial^2D{\hat u}+2{\hat u}
\partial^2D+{\hat\alpha}\partial^2+\partial^2{\hat \alpha}
+2{\hat v}\partial D+2{\hat u}\partial D{\hat u}
-{\hat u}\partial{\hat\alpha}\\
&+{\hat\alpha}\partial {\hat u}+\partial (D{\hat v})
+{\hat v}_xD-{\hat u}(D{\hat v})+
{\hat v}(D{\hat u})-2{\hat v}{\hat \alpha}-{\hat\beta}_x+
2{\hat\beta}{\hat u},\\
{\hat P}_{34}=&\partial^4+\partial^3{\hat u}-{\hat u}\partial^3
-\partial^2D{\hat\alpha}-{\hat\alpha}\partial^2D-
{\hat u}\partial^2{\hat u}-\partial^2{\hat v}+
{\hat u}\partial D {\hat\alpha}-{\hat\beta}\partial D
\\
&-{\hat\alpha}\partial D{\hat u }+{\hat\alpha}\partial
{\hat\alpha}+{\hat u}\partial{\hat v}+{\hat\alpha}D{\hat v}
-{\hat\beta}D{\hat u}-
2{\hat\alpha}{\hat\beta},\\
{\hat P}_{41}=&2\partial^3-2{\hat u}\partial^2-2{\hat\alpha}
\partial D-2{\hat v}\partial-D{\hat \beta},
\hspace*{2cm}
{\hat P}_{42}=-{\hat\beta}_x-3{\hat\beta}\partial,\\
{\hat P}_{43}=&-\partial^4-\partial^3{\hat u}+{\hat u}\partial^3
+\partial^2 D{\hat\alpha}+{\hat\alpha}\partial^2 D+
{\hat u}\partial^2{\hat u}
+{\hat v}\partial^2-{\hat u}\partial D{\hat\alpha}
+{\hat \alpha}\partial D{\hat u}\\
&-\partial D{\hat\beta}-{\hat\alpha}\partial {\hat\alpha}
-{\hat v}D{\hat\alpha}+{\hat v}\partial{\hat u}-2{\hat \beta}{\hat\alpha}
+{\hat u}D{\hat\beta},\\
{\hat P}_{44}=& {\hat u}\partial {\hat\beta}+
{\hat\beta}\partial {\hat u}
-2{\hat\beta}_x\partial+{\hat\alpha}D{\hat\beta}-{\hat\beta}D
{\hat \alpha}-{\hat\beta}_{xx},
\end{align*}
\end{small}

\vspace*{-1cm}

\hspace*{-.6cm}\rule{1\textwidth}{0.15 mm}\\ 
\vspace*{-.94cm}
   
\hspace*{-.6cm}\rule{1\textwidth}{0.2 mm}   

\bigskip

Through the relationship between $L_{2}^{(2)}$ and $L$,
we have a change of coordinates
\begin{equation}\label{m1}
{\hat u}=-u,\quad
{\hat\alpha}=\alpha-(Du),\quad
{\hat v}=-v-(D\alpha),\quad
{\hat\beta}=-\beta-(Dv),
\end{equation} 
this is obviously an invertible change of variables.
By laborious but straightforward computation, one can check
that the above map is a Hamiltonian
map between $P$ and ${\hat P}$.

It is known that suitable combinations of hatted variables lead
to primary fields, namely
\begin{equation}\label{m2}
\begin{aligned}
&J={\hat u},\\
&T={\hat\alpha}-{1\over 2}(D{\hat u}),\\
&W_2={\hat v}-{1\over 3}(D{\hat\alpha})-{1\over 3}{\hat u}_x+{2\over 9}{\hat 
u}^2,\\
&W_{5\over 2}={\hat\beta}-{1\over 2}(D{\hat v})
-{1\over 2}{\hat\alpha}_x+{1\over 6}(D{\hat u})_x+{4\over 9}{\hat u}{\hat\alpha}
-{2\over 9}{\hat u}(D{\hat u}),
\end{aligned}
\end{equation}
thus the composition of (\ref{m1}) and (\ref{m2}) will give us
the correct formulation for unhatted variables.

Above argument reveals that
the hierarchy we constructed from $L_{2}^{(2)}$ is a hierarchy associated 
with $N=2$ super $W_3$ algebra. We notice that
these hierarchies were constructed by Yung\cite{y} and
Pichugin {\em et al} \cite{bi} independently. Those authors
presented three possible integrable $N=2$ supersymmetric
Boussinesq systems. It is further shown one of these 
hierarchies has two local Hamiltonian operators\cite{p}\cite{bi}.  
Our hierarchy is equivalent to this system. To see this, 
we notice that the first Hamiltonian structure
for this hierarchy is obtained by a simple shift
of the field $W_2$, which corresponds a shift of $v$
in our coordinates. However this simple shift
precisely gives us the first Poisson tensor $Q$.
Since two compatible Hamiltonian
operators determines a hierarchy uniquely, we conclude
that these hierarchies are indeed same. Another point we
can now make is the non-reducible of the extended Boussinesq
system to
the classical Boussinesq system. This should be clear
from the form our Lax operator.

\smallskip

\noindent{\bf Acknowledgment}

I should like to thank the anonymous referee for his critical comments
which has led to improvement of the paper.


\begin{thebibliography}{99}
\bibitem{bi} S. Bellucci, E. Ivanov, S. Krivonos and
  A. Pichugin, {\em Phys. Lett. B} {\bf 312} (1993) 463.
\bibitem{b} L. Bonora, S. Krivonos and A. Sorin,
{\em Nucl. Phys. B} {\bf 477} (1996) 835.
\bibitem{bd} J.C. Brunelli and A. Das, {\em Phys.Lett. B} {\bf 337}
(1994) 303; {\bf 354} (1995) 307.
\bibitem {fr} J.M. Figueroa-O'Farrill and E. Ramos,
{\em Nucl. Phys. B} {\bf 368} (1992) 361.
\bibitem{hn} K. Huitu and D. Nemeschansky, {\em Mod. Phys. Lett. A}
{\bf 6} (1991) 3179.
\bibitem{ik} T. Inami and H. Kanno, {\em J. Phys. A: Math. Gen} {\bf 25}
(1992) 3729.
\bibitem{ik1} T. Inami and H. Kanno, {\em Int. J. Mod. Phys. A} {\bf 7} Suppl. 
1A (1992) 419.
\bibitem{lm} C.A. Laberge and P. Mathieu, {\em Phys. Lett. B}
{\bf 215} (1988) 718.
\bibitem{mp} C. Morosi and L. Pizzocchero, {\em Commun. Math. Phys.}
{\bf 158} (1993) 267;
{\em Phys. Lett. A} {\bf 185} (1994) 241.
\bibitem{o} W. Oevel, {\em Phys. Lett. A} {\bf 186} (1994) 79.

\bibitem{os} W. Oevel and W. Strampp, {\em Commun. Math. Phys.}
{\bf 157} (1993) 51.

\bibitem{p} Z. Popowicz, {\em Phys. Lett. B} {\bf 319} (1993)
487.

\bibitem{y} C.M. Yung, {\em Phys. Lett. B} {\bf 309} (1993) 75.
\bibitem{yw} C.M. Yung and R.C. Warner, {\em J. Math. Phys.}
{\bf 34} (1993) 4050.



\end{thebibliography}
\end{document}